\documentclass[aps,prb,twocolumn,groupedaddress,showpacs,floatfix]{revtex4}
\usepackage{graphicx}
\begin{document}
\pagenumbering{arabic}
\pagestyle{plain}
\title{Proton NMR measurements of the local magnetic field in the paramagnetic metal and antiferromagnetic insulator phases of $\lambda$-(BETS)$_{2}$FeCl$_{4}$}  
\author{Guoqing Wu,$^{1}$ P. Ranin,$^{1}$ W. G. Clark,$^{1}$ S. E. Brown,$^{1}$ L. Balicas,$^{2}$ and L. K. Montgomery$^{3}$}  
\affiliation{$^{1}$Department of Physics and Astronomy, UCLA, Los Angeles, California 90095-1547, USA}
\affiliation{$^{2}$National High Magnetic Field Laboratory, Florida State University, Tallahassee, Florida 32306, USA}
\affiliation{$^{3}$Department of Chemistry, Indiana University, Bloomington, IN 47405, USA}
\date{\today}
\begin{abstract}
   Measurements of the $^{1}$H-NMR spectrum of a small ($\sim$ 4 $\mu$g) single crystal of the organic conductor $\lambda$-(BETS)$_{2}$FeCl$_{4}$ are reported with an applied magnetic field $\bf{B}$$_{0}$ = 9 T parallel to the $a$-axis in the $ac$-plane over a temperature $(T)$ range 2.0 $-$ 180 K. They provide the distribution of the static local magnetic field at the proton sites in the paramagnetic metal (PM) and antiferromagnetic insulator (AFI) phases, along with the changes that occur at the PM$-$AFI phase transition. The spectra have six main peaks that are significantly broadened and shifted at low $T$. The origin of these features is attributed to the large dipolar field from the 3d Fe$^{3+}$ ion moments (spin $S_{\rm{d}}$ = 5/2). Their amplitude and $T-$dependence are modeled using a modified Brillouin function that includes a mean field approximation for the total exchange interaction ($J_{0}$) between one Fe$^{3+}$ ion and its two nearest neighbors. A good fit is obtained using $J_{0}$ = $-$ 1.7 K. At temperatures below the PM$-$AFI transition temperature $T_{MI}$ = 3.5 K, an extra peak appears on the high frequency side of the spectrum and the details of the spectrum become smeared. Also, the rms linewidth and the frequency shift of the spectral distribution are discontinuous, consistent with the transition being first-order. These measurements verify that the dominant local magnetic field contribution is from the Fe$^{3+}$ ions and indicate that there is a significant change in the static local magnetic field distribution at the proton sites on traversing the PM to AFI phase transition. 
\end{abstract}
\pacs{75.30.Kz, 75.50.Ee, 76.60.-k, 71.30.+h}
\maketitle
\section{Introduction}
    The organic conductor $\lambda$-(BETS)$_{2}$FeCl$_{4}$, where BETS is bis(ethylenedithio)tetraselenafulvalene (C$_{10}$S$_{4}$Se$_{4}$H$_{8}$), is of considerable interest because of the properties related to the coexistence of the large magnetic 3d Fe$^{3+}$ moments (spin $S_{\rm{d}}$=5/2) of the inorganic anions (FeCl$^{4-}$) and the conduction $\pi$-electrons (spin $S_{\pi}$=1/2) in the donor molecules from the BETS.\cite{uji1, uji2, kobayashi1, tokumoto, brossard, akutsu1} It has an unusual phase diagram, including an antiferromagnetic insulating (AFI) phase, a paramagnetic metallic (PM) phase, and a field-induced superconducting (FISC) phase.\cite{uji1, uji2} Also, results interpreted in terms of a ferroelectric phase transition in the metallic phase\cite{endo, negishi} at 70 K and a relaxor ferrolectric behavior\cite{matsui} at 30 K have been reported.

    A mechanism used to explain the FISC phase below 5 K in $\lambda$-(BETS)$_{2}$FeCl$_{4}$ is based upon the Jaccarino-Peter (J-P) compensation effect\cite{jaccarino} operating in a two-dimensional (2D) system.\cite{uji1, uji2, balicas1, balicas2} In this model, the negative exchange interaction ($J_{\pi\rm{d}}$) between the paramagnetic 3d Fe$^{3+}$ moment ($g\mu_{B}S_{\rm{d}}$) ($g$ is the Land\'{e} factor and $\mu_{B}$ is the Bohr magneton) and the conduction $\pi$-electrons in the BETS molecule ($\pi-$d interaction)\cite{uji1, akutsu1} generates a large magnetic field ($\bf{B}$$_{\pi\mathrm{d}}$) that cancels most of the externally applied magnetic field ($\bf{B}$$_{0}$) when the latter is large ($B_{0}$ $\sim$ 17 $-$ 45 T) and aligned parallel to the $ac$ plane. This suggests that $\mathbf{B}$$_{\pi\rm{d}}$ ($B_{\pi\rm{d}}$ = $J_{\pi\rm{d}}<S_{\rm{d}}>$/$g\mu_{B}$, where $<S_{\rm{d}}>$ is the average value of the Fe$^{3+}$ spin polarization) at the conducting $\pi$-electrons is on the order of $\sim$ 30 T and its direction is antiparallel to $\bf{B}$$_{0}$. However, this interpretation has been challenged because the superconducting state in $\lambda$-(BETS)$_{2}$FeCl$_{4}$ can be destroyed by a very small amount of out-of-plane magnetic field ($\sim$ 0.1 T)\cite{uji1, uji2} and the alignment of the paramagnetic Fe$^{3+}$ moment follows closely that of $\bf{B}$$_{0}$. It has also been proposed that the Larkin-Ovchinkov-Fulde-Farrell (LOFF) phase is present near the boundary of the FISC phase.\cite{fulde, houzet}

    The magnetic PM$-$AFI phase transition, which occurs at $B_{0}$ $<$ 11 T (N$\acute{\rm{e}}$el temperature $T_{N}$), coincides with a metal-insulator (MI) transition\cite{uji1, akutsu2, akutsu3} (transition temperature $T_{MI}$). The property $T_{N}$ = $T_{MI}$ indicates that the MI and AFI transitions are cooperative transitions.\cite{uji1, akutsu2, akutsu3} Thus, it is expected this PM$-$AFI transition is also a result of the $\pi-$d interaction, since the study of its iso-structural nonmagnetic and non-3d-electron analog $\lambda$-(BETS)$_{2}$GaCl$_{4}$ shows that it exhibits a behavior \cite{kobayashi2, kobayashi3} that is completely different from that of $\lambda$-(BETS)$_{2}$FeCl$_{4}$. However, the detailed role of the $\pi-$d interaction for the PM$-$AFI phase transition is not yet clear because it has not yet been established by direct experimental evidence.

    Prior reports of proton NMR measurements on $\lambda$-(BETS)$_{2}$FeCl$_{4}$ include the spectra of a large ($\sim$ 6.5 mg) aggregate of crystals aligned along the $c$-axis in a magnetic field \cite{endo} of 2.2 T and a preliminary report \cite{wgc} of our measurements on a $\sim$ 4 $\mu$g single crystal, for which the spectrum results are presented and analyzed in detail here. A comparison of the results from the aggregate sample\cite{endo} and the significantly different ones on the single crystal reported here is discussed is Section IV.F.
\begin{figure}
\includegraphics[scale= 0.43]{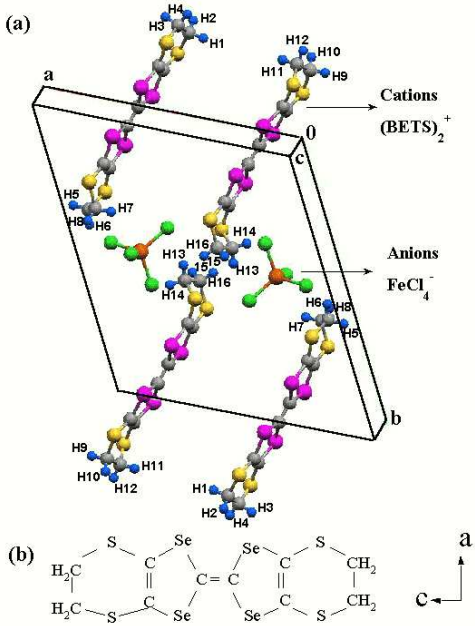}
\caption{(color online) (a) The crystal structure of $\lambda$-(BETS)$_{2}$FeCl$_{4}$ with all 32 hydrogen atoms labeled in a unit cell. Atoms in color: Red $-$Fe; Green $-$Cl; Blue $-$H; Grey $-$C; Yellow $-$S; Pink $-$Se. (b) The BETS molecule, which identifies the Se, C, and S atoms. \label{fig1}}
\end{figure}

    The crystal structure\cite{kobayashi1, kobayashi2} of $\lambda$-(BETS)$_{2}$FeCl$_{4}$, shown in Fig. 1, is triclinic with the space group P$\bar{1}$ and the lattice constants at 298 K $a$ = 16.164(3) $\rm{\AA}$, $b$ = 18.538(3) $\rm{\AA}$, $c$ = 6.5928(8) $\rm{\AA}$, $\alpha$ = 98.40(1)$^{\circ}$, $\beta$ = 96.67(1)$^{\circ}$, $\gamma$ = 112.52(1)$^{\circ}$ and $V$ = 1773.0(5) $\rm{\AA}^{3}$. There are four BETS molecules and two Fe$^{3+}$ ions per unit cell and the BETS moleIcules are stacked along the $a$- and $c$-axes to form a quasi-stacking fourfold structure. The conducting layers, comprised of BETS, are sandwiched along the $b$-axis by the insulating layers of FeCl$^{4-}$ anions. The least conducting axis is $b$, $ac$ is the conducting plane, and the easy axis of the antiferromagnetic spin structure is $\sim$ 30$^{\circ}$ away from the $c$ axis ($||$ needle axis of the crystal).\cite{hk}
    
    In this paper, measurements of the proton NMR spectrum of a single $\sim$ 4 $\mu$g crystal of $\lambda$-(BETS)$_{2}$FeCl$_{4}$ are reported with a magnetic field $\bf{B}$$_{0}$ = 9 T applied parallel to the $a$-axis in the $ac$-plane over a temperature ($T$) range 2.0$-$180 K. Along with our preliminary report, it is the first published report of proton NMR in a single crystal of this material. It is a challanging experiment involving an extremely small sample size ($\sim$ 0.018 mm $\times$ 0.065 mm $\times$ 1.2 mm). These measurements probe the distribution and the origin of the static local magnetic field at the proton sites in the PM and AFI states as well as across the PM$-$AFI phase transition. The observed properties should help to establish a microscopic model for the PM$-$AFI phase transition.

    One important result of this investigation is that the dominant local magnetic field at the proton sites comes from the large dipolar field of the 3d Fe$^{3+}$ ion moments. A mean field model based on the dipolar field of the Fe$^{3+}$ moments is presented and used to calculate the proton NMR spectrum. It provides a good fit to the measured spectra. Besides this, the total exchange constant $J_{0}$ between an Fe$^{3+}$ ion and its two nearest neighbors is determined to be $J_{0}$ $\sim$ $-$1.7 K from a fit to the spectrum data. These measurements also show that there is a significant change in the static local magnetic field distribution at the proton sites across the PM$-$AFI phase transition. No proton NMR evidence of a ferroelectric phase transition at 70 K is observed in these measurements.

    The rest of this paper is organized as follows. Section II presents the experimental details and Section III has the experimental results for the proton NMR spectra, including frequency distributions, shifts, and linewidths. Section IV presents the model for the spectrum, along with the comparison with the measured spectra. The conclusions are stated in Section V.

\section{experimental details}

    The needle-like single crystal $\lambda$-(BETS)$_{2}$FeCl$_{4}$ samples were prepared as described by Montgomery et al. with a standard electrochemical oxidation method.\cite{montgomery} The dimensions of the sample used for these $^{1}$H-NMR measurements are (1.2 $\pm$ 0.1) mm $\times$ (0.065 $\pm$ 0.010) mm $\times$ (0.018 $\pm$ 0.005) mm, which corresponds to (3.8 $\pm$ 1.8) $\mu$g in mass and (2.7 $\pm$ 1.3) $\times$ 10$^{16}$ protons.

    The NMR coil used was 40 turns of 0.025 mm diameter bare copper wire wound on a 0.075 mm diameter wire form. The coil was held to the rest of the probe circuit by two 125 $\mu$m diameter Cu wire leads with Teflon insulation, which was removed close to the ends where the coil was soldered to them. Commercial pure acetone was used for cleaning the coil and its surroundings when the NMR coil was set on the probe to reduce the spurious proton signals relative to the signals from such a small sample.

    Finally, a single piece of the needle shape $\lambda$-(BETS)$_{2}$FeCl$_{4}$ single crystal sample was slid into the coil, carefully aligned close to $B_{0}$ $\parallel$ $a$ in the $ac$-plane, and held in place with a very small amount of commercial Apiezon grease on each end. The orientation is done visually under a microscope with an estimated uncertainty of $\sim$ $\pm$ 5$^{\circ}$. [Note: the $ac$-plane is $\sim$ in the sample surface plane which has the largest surface area, the $c$-axis $||$ the needle direction of the sample, and the angle between the $a$- and $c$-axes is $\beta$ ($\beta$ = 96.67(1$^{\circ}$) at 298 K)].\cite{kobayashi1, kobayashi2, montgomery}

    As shown in the preliminary report,\cite{wgc} the spurious proton signal was estimated to be less than $\sim$ 4$\%$ of the signal from the $\lambda$-(BETS)$_{2}$FeCl$_{4}$ sample by comparing the signal with and without the sample in the coil. Thus, in these measurements the spurious proton signal has an insignificant size.

    The $^{1}$H-NMR frequency-swept spectra were obtained using standard spin-echo techniques carried out with a spectrometer and probe built at UCLA. Since the proton has a gyromagnetic ratio $\gamma_{I}$ = 42.5759 MHz/T, the frequency $\nu$ for the excitation pulses used for the spectrometer is near $\nu$ $\sim$ $\nu_{0}$ = $\gamma_{I}B_{0}$ = 382.6935 (MHz), where $\nu_{0}$ is the proton Larmor frequency in the external field. The value of $B_{0}$ used in this experiment was $B_{0}$ = 8.9885 T (for simplicity, often referred to here as 9 T).

    Because the NMR spectrum covers a wide range in frequency up to 14 MHz (3.3 kG), short rf pulses and a wide receiver bandwidth ($\pm$ 1 MHz) were used to record the spin-echo signals. The pulse sequence that optimized the height of the spin echo used to record the NMR signal was a 0.2 $\mu$s $\pi$/2 pulse ($B_{1}$ = 294 G, 1.25 MHz proton frequency) followed by a 0.3 $\mu$s pulse separated by a time interval $\tau$ ($\tau$ $\sim$ 5 $\mu$s) for most of the measurements. For a viable signal-to-noise ratio, each echo signal was averaged 2000 times at 180 K and 128 times at 4.2 K and lower temperatures. The uncertainty associated with the signal-to-noise ratio is probably the main source of error in the data. The uncertainties include $\sim$ $\pm$ 1$\%$ in $T$ and $\pm$ 5$^{\circ}$ in the field alignment.

    At low $T$, the spectrum is very wide ($\sim$ 12 MHz) and the frequency sweep covered a range as high as 370 to 400 MHz and used a typical frequency step for each acquisition of 0.2 $-$ 0.5 MHz. When a wide frequency sweep range was used, the probe circuit was retuned every 4 MHz to maintain a uniformly high sensitivity (above 85$\%$) for recording the proton spectrum. The spectra were analyzed with frequency-shifted and -summed Fourier transform processing.\cite{clark}
\section{results}
\subsection{$^{1}$H-NMR spectra}
\begin{figure}
\includegraphics[scale= 0.64]{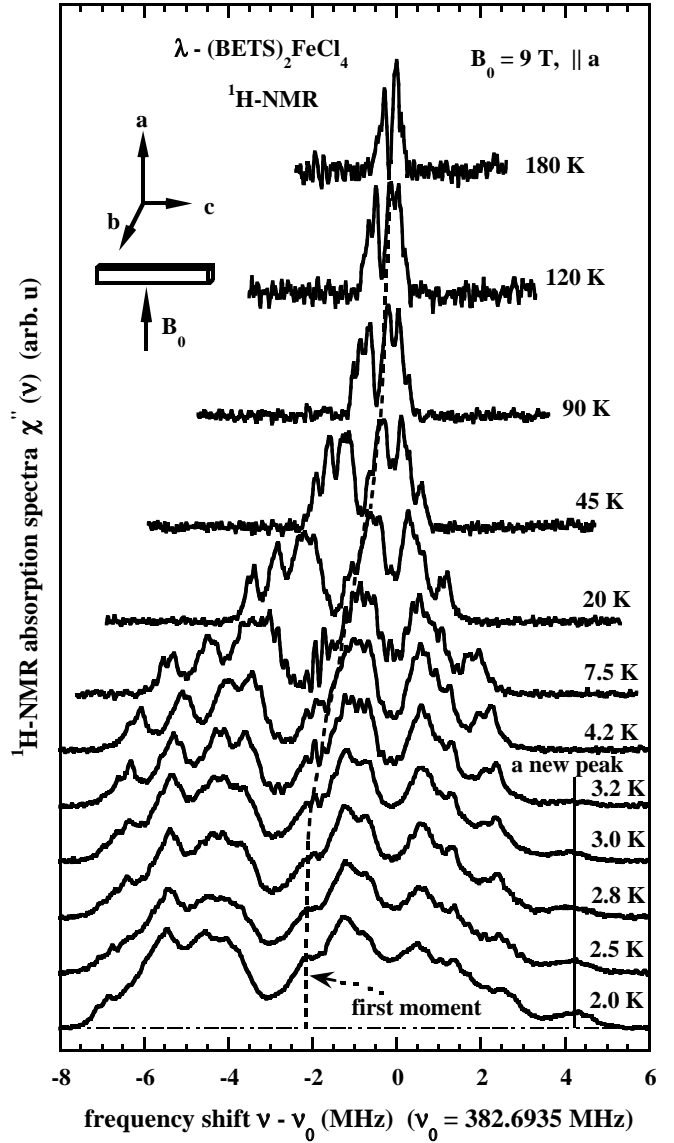} 
\caption{Normalized $^{1}$H-NMR absorption spectra of a single crystal of $\lambda$-(BETS)$_{2}$FeCl$_{4}$ as a function of $\nu~ - ~\nu_{0}$ ($\nu_{0}$ = 382.6935 MHz) from 180 K to 2 K with $\bf{B}$$_{0}$ = 8.9885 T parallel to the $a$-axis in the $ac$-plane. The solid vertical line at $\nu~ - ~\nu_{0}$ = 4.25 MHz indicates a new peak below $T$ $\leq$ $T_{MI}$ = 3.5 K, and the dashed line shows $<\nu>$ for each spectrum. \label{fig2}}
\end{figure}
    Figure 2 shows the normalized $^{1}$H-NMR absorption spectra $[\chi^{''}(\nu)]$ of a $\lambda$-(BETS)$_{2}$FeCl$_{4}$ single crystal as a function of the frequency shift $\nu$ $-$ $\nu_{0}$ with $\bf{B}$$_{0}$ = 8.9885 T parallel to the $a$-axis in the $ac$-plane for $T$ = 180 K to 2 K. 

    Over most of this range, the spectra have six main peaks which can be divided into two groups (low frequency side and high frequency side) with 3 main peaks for each. As $T$ is lowered from 180 to 4.2 K in the PM state, the spectrum broadens significantly, its center shifts to lower frequency and the splitting between the peaks increases. For such a complex spectrum, a reasonable measure of its center is the first moment, $<\nu>$, given by \cite{slichter}
\begin{equation}
<\nu>~=~\frac{\sum_{i}\nu_{i}\chi_{i}^{''}(\nu_{i})}{\sum_{i}\chi_{i}^{''}(\nu_{i})},\\
\end{equation}
where $i$ indexes equally spaced frequency steps. The average shift of the spectrum ($\Delta\nu$) indicated by the dashed line in Fig. 2, is $\Delta \nu = <\nu> - ~\nu_{0}$. Also, some additional weak structures gradually develop at lower $T$. As the sample is cooled further into the AFI phase ($T$ $\leq$ $T_{MI}$ = 3.5 K), the details of the spectrum become somewhat smeared, an additional peak at 4.25 MHz appears and it grows larger with further cooling (solid line in Fig. 2).

    As discussed in later sections, this relatively complex spectrum is caused mainly by the dipolar field of the 3d Fe$^{3+}$ ion electron spin moments ($S_{d}$ = 5/2, $g~ \approx$ 2) at 16 magnetically inequivalent proton sites in both the PM and AFI phases. In both phases, these moments are present, the major differences being that they should have long-range order and a different orientation in the AFI state.

    At the lowest $T$ in the PM state, the average mganetization of each Fe$^{3+}$ ion is almost completely saturated by $B_{0}$. Therefore, the main changes in the NMR spectrum associated with the AFI phase are caused by changes in the orientation of the Fe$^{3+}$ average spin moments, not by changes in their magnitude.
    
    In the AFI phase ($T$ $\leq$ $T_{MI}$), the changes of the spectrum vs $T$, such as the smearing of the spectrum details and the growth of the new peak at the high frequency side, occur gradually rather than suddenly. This behavior indicates a continuous change of the local magnetic field distribution at the proton sites as $T$ is lowered. Its origin is probably the development of the canted-antiferromagnetic phase as observed from the field dependence of magnetization and capacitance measurements reported\cite{uji2, brossard} for $\lambda$-(BETS)$_{2}$FeCl$_{4}$.
\subsection{$^{1}$H-NMR frequency shift}
     Figure 3 shows the frequency shift magnitude $|\Delta\nu|$ as a function of $T$, and the inset of Fig. 3 plots 1/$|\Delta\nu|$ vs T. A clear discontinuity of $\Delta\nu$ at $T$ = $T_{MI}$ is seen. The error bars are our best estimate of the uncertainty in our data analysis.

     The NMR frequency at the $i$th proton site is $\gamma_{I}\times|\bf{B}$$_{i}|$ , where $\bf{B}$$_{i}$ is the total magnetic field at the site. Since $\bf{B}$$_{i}$ is the sum of the large applied $\bf{B}$$_{0}$ plus the much smaller field generated by the sample ($\Delta\bf{B}$$_{0}$), it is easily shown that the field shift caused by the magnetic properties of the sample is the component of its field parallel to $\bf{B}$$_{0}$ ($\Delta B_{||}$). The values of $\Delta\nu$ in Fig. 3 represent the average of $\Delta B_{||}$ over the proton sites ($<\Delta B_{||}>$) with
\begin{equation}
\Delta\nu=\gamma_{I}<\Delta B_{||}>.\\
\end{equation}
Thus, the origin of $\Delta\nu$ is the distribution of $\Delta B_{||}$ over the proton sites.
\begin{figure}
\includegraphics[scale= 0.42]{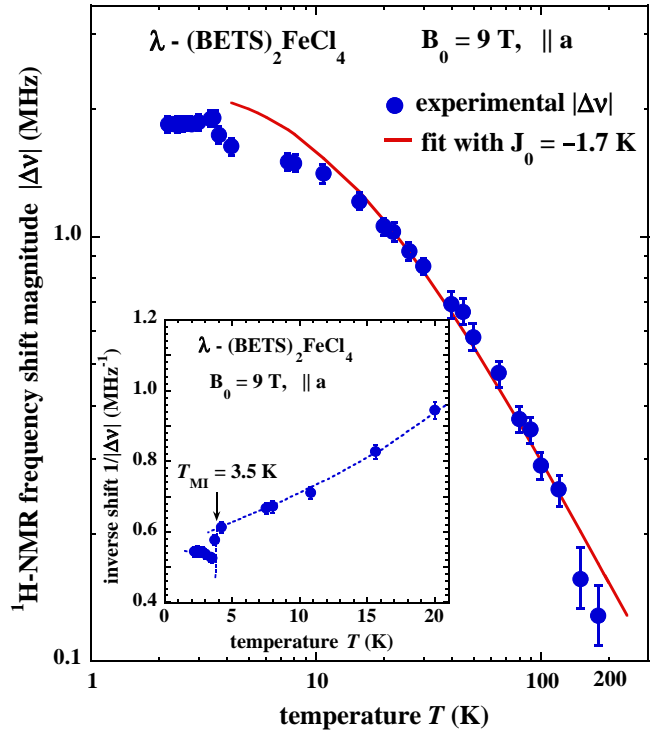} 
\caption{(color online) $T-$dependence of $|\Delta\nu|$ for a single crystal of $\lambda$-(BETS)$_{2}$FeCl$_{4}$ from 180 K to 2 K with $\bf{B}$$_{0}$ = 8.9885 T parallel to the $a$-axis in the $ac$-plane. The solid (red) line is a fit to Eqs. (6)$-$(10) with $J_{0}$ = $-$ 1.7 K. The inset shows the discontinuity of 1/$|\Delta\nu|$ vs $T$ across the PM$-$AFI phase transition at 3.5 K. The dashed (blue) line is a guide to the eye. \label{fig3}}
\end{figure}

     At 180 K, $\Delta\nu$ has a value of $-$(0.13 $\pm$ 0.02) MHz, while at 4 K it reaches $-$(1.750 $\pm$ 0.005) MHz. Below $T_{MI}$ = 3.5 K, $\Delta\nu$ has a rather weak temperature dependence, as shown in the inset of Fig. 3. The sudden decrease of 1/$|\Delta\nu|$ at $T_{MI}$ indicates an increase of the average static local magnetic field at the proton sites due to the PM$-$AFI phase transition. Its sharpness is evidence that the PM$-$AFI phase transition is first order.

     The negative sign of $\Delta\nu$ indicates that the direction of the average static local magnetic field is opposite to the external magnetic field. But this does not mean that the average static moment is negative. The negative sign is determined by the proton positions relative to the Fe$^{3+}$ ions in the crystal lattice.
\subsection{$^{1}$H-NMR spectrum linewidth}
     A reasonable quantitative measure of the width of the local field distribution that generates the proton spectrum of $\lambda$-(BETS)$_{2}$FeCl$_{4}$ is the root mean square (rms) linewidth, $\Delta f_{\rm{rms}}$, i.e. the square root of the second moment, $<\Delta\nu^{2}>^{1/2}$, given by \cite{slichter, abragam}
\begin{eqnarray}
\Delta f_{\rm{rms}} \equiv <\Delta\nu^{2}>^{1/2} = \left[\frac{\sum_{i}(\nu_{i}-<\nu>)^{2}\chi^{''}(\nu_{i})}{\sum_{i}\chi^{''}(\nu_{i})}\right]^{1/2}.\\
\nonumber
\end{eqnarray}
\begin{figure}
\includegraphics[scale= 0.42]{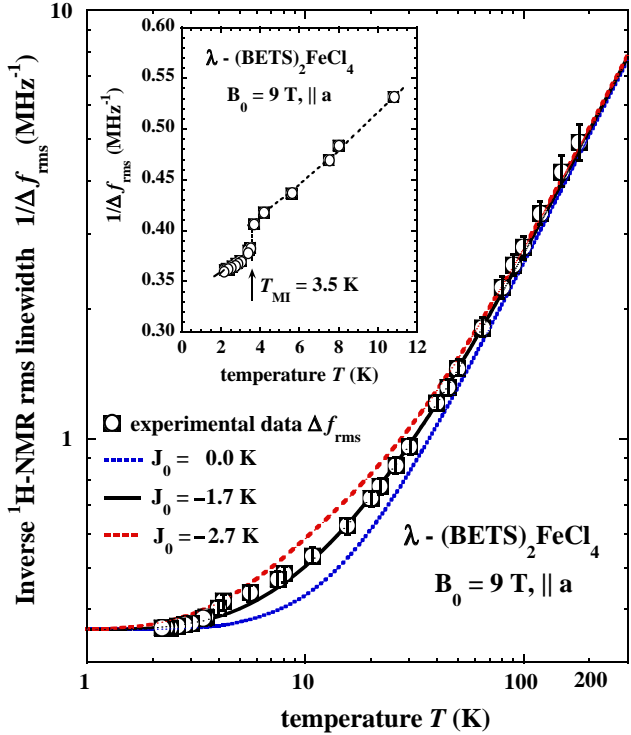}
\caption{(color online) $T-$dependence of 1/$\Delta f_{\rm{rms}}$ for a single crystal of $\lambda$-(BETS)$_{2}$FeCl$_{4}$ from 180 K to 2 K with $\bf{B}$$_{0}$ = 8.9885 T parallel to the $\it{a}$-axis in the $ac$-plane. The solid (black) line shows the best fit to the data from Eqs. (6)$-$(10) with $J_{0}$ = $-$ 1.7 K. The inset shows the low $T$ data across the PM$-$AFI phase transition, with a dashed line as a guide to the eye. \label{fig4}}
\end{figure}

     The measurements of $\Delta f_{\rm{rms}}$ for protons as a function of $T $ in $\lambda$-(BETS)$_{2}$FeCl$_{4}$ are shown in Fig. 4, where 1/$\Delta f_{\rm{rms}}$ is plotted as a function of $T$. At 180 K, $\Delta f_{\rm{rms}}$ has the value of (0.205 $\pm$ 0.005) MHz and it reaches (2.50 $\pm$ 0.10) MHz at $T_{MI}$ = 3.5 K. The error bar is estimated to be equal to or smaller than the size of the marker below $\sim$ 50 K.

     Across $T_{MI}$ = 3.5 K, as for $|\Delta\nu|$ in Fig. 3, there is a sudden increase of $\Delta f_{\rm{rms}}$ in Fig. 4 that also indicates a first order phase transition.

     There are several important properties of the data shown in Figs. 2$-$4. At $T$ $>$ 50 K, it can be shown that the $T-$dependence of all of them ($<\nu>$, $|\Delta\nu|$, and $\Delta f_{\rm{rms}}$) follows the Curie-Weiss relation with a Curie-Weiss temperature $\Theta\simeq$ 5.5 K. They also follow the Brillouin function $[B_{J}(x)]$ behavior using the parameters for the Fe$^{3+}$ spins. This result is a strong indication that the dominant contribution to $\Delta \bf{B}$$_{0}$ is the dipolar field at the proton sites from the magnetization of the Fe$^{3+}$ spins.

     An important issue for characterizing the data is that they extend well into the low $T$ regime; i.e., at 9 T, the Zeeman splitting between the highest and lowest Fe$^{3+}$ spin states is 5 $\times~ g\mu_{\rm{B}}B_{0}/k_{\rm{B}}$ $\simeq$ 60.5 K, where $k_{\rm{B}}$ is the Boltzmann constant. For this reason, instead of the more commonly used Curie$-$Weiss law, which is a high $T$ approximation ($T >> g\mu_{\rm{B}}S_{\rm{d}}/k_{\rm{B}}$ $\approx$ 30.2 K), we characterize the data with the Brillouin function, which includes magnetic saturation at low $T$. Also, since there is a significant total antiferromagnetic (negative) exchange interaction $J_{0}$ among the Fe$^{3+}$ moments, \cite{brossard, mori} the Brillouin function used here is modified to include a simple, approximate mean field correction.

     For non-interacting moments, the magnetization $[M(x_{0})]$ is given by \cite{ashcroft}
\begin{equation}
M(x_{0}) = N_{\rm{A}}g\mu_{\rm{B}}J~ B_{J}(x_{0}),\\
\end{equation}
where $N_{\rm{A}}$ is Avogadro's number, $J$ = $S_{\rm{d}}$ = 5/2 for the Fe$^{3+}$, and $B_{J}(x_{0})$ is
\begin{equation}
B_{J}(x_{0}) = \frac{2J + 1}{2J}\coth\left( \frac{2J + 1}{2J}x_{0}\right) - \frac{1}{2J}\coth\left( \frac{1}{2J}x_{0}\right), \\
\end{equation}
where
\begin{equation}
x_{0} = \frac{Jg\mu_{\rm{B}} B_{0}}{k_{\rm{B}}T}.\\
\end{equation}
The effect of $J_{0}$ for an antiferromagnetic exchange interaction between the nearest neighbor Fe$^{3+}$ moments can be modeled as an additional magnetic field component ($B^{\prime}$) antiparallel to $\bf{B}$$_{0}$,\cite{ashcroft} given by
\begin{eqnarray}
B' = \frac{|J_{0}|Jk_{\rm{B}}}{g\mu_{\rm{B}}}B_{J}(x) \approx \frac{|J_{0}|Jk_{\rm{B}}}{g\mu_{\rm{B}}}B_{J}(x_{0}), \\
\nonumber
\end{eqnarray}
where 
\begin{eqnarray}
  x = x_0 + x' = \frac{Jg\mu_{\rm{B}}(B_{0} - B')}{k_{\rm{B}}T},  \\
\nonumber
\end{eqnarray}
and the right side of Eq. 7 has been used for $B^{\prime}$. Because the value obtained later for $B^{\prime}$ is substantially smaller than $B_{0}$, this approximation is a reasonable one. Thus, the mean field modified Brillouin function used to model our data is $B_{J\rm{M}}(x)$ and the corresponding formula to fit $\Delta f_{\rm{rms}}$ is
\begin{eqnarray}
\Delta f_{\rm{rms}} = C' ~M(x) = C'N_{\rm{A}}g\mu_{\rm{B}}J~B_{J\rm{M}}(x),  \\
\nonumber
\end{eqnarray}
where 
\begin{equation}
B_{J\rm{M}}(x) = B_J(x_0 + x'), \\
\end{equation}
and $C'$ and $J_0$ are adjusted to give the best fit to the data.

     As shown in Fig. 4, from the $T-$dependence of $\Delta f_{\rm{rms}}$ the best fit is obtained with $J_{0}$ = $-$(1.7 $\pm$ 0.2) K. This corresponds to a maxmium total exchange field of $\sim$ $-$ 3 T below $\sim$ 5 K. The negative sign of $J_{0}$ indicates that the Fe$^{3+}$ ions have an antiferromagnetic (AF) exchange interactions with their nearest neighbors. Since each Fe$^{3+}$ ion has two nearest neighbors \cite{kobayashi1} (actually three closest ones: two are at $\sim$ 6.6 $\AA$ away, and one at $\sim$ 7.6 $\AA$ away) the exchange constant $J_{\rm{dd}}$ between each pair of Fe$^{3+}$ ions is, $J_{\rm{dd}}$ $\approx$ $J_{0}$/2 = $-$(0.85 $\pm$ 0.10) K, which agrees with theoretical expectations ($\sim-$ 0.64 K).\cite{mori} The parameter $C^{\prime}$ has a fitted value of (100 $\pm$ 6) $[$(mol.Fe/emu)Hz$]$ [Note: The units for $M(x)$ are emu/mol.Fe; see Eqs. (4) and (9)]. The overall difference between the fit and the $\Delta f_{\rm{rms}}$ data is below $\sim$ 5$\%$, except near the phase transition, where it is $\sim$ 10$\%$.

     Similarly, $|\Delta\nu|$ (Fig. 3) is well characterized by Eqs. (6)$-$(10) from 180 K down to 10 K with the same fit value of $J_{0}$. But its deviation is slighly larger below $\sim$ 10 K. This is possibly caused by not including the demagnitization and Lorentz fields to the local field $B^{\prime}$ in Eq. (7) for the shift.

     The property $\Delta f_{\rm{rms}}$ $>$ $|\Delta\nu|$ shows that there is a broad static local magnetic field distribution in $\lambda$-(BETS)$_{2}$FeCl$_{4}$. It occurs because there are 16 inequivalent $^{1}$H-sites at which the dipolar field from the Fe$^{3+}$ moments has a large variation.

     As discussed in more detail below, these proton shift properties support the conclusion that they are dominated by the dipolar field from the 3d Fe$^{3+}$ ion electron moments. The sudden change in the spectrum at $T_{MI}$ = 3.5 K, as well as seen from those proton shift properties, reflects a comprehensive change of the static local magnetic field distribution at the $^{1}$H-sites due to the AF ordering of the Fe$^{3+}$ electron spins. 
\section{discussion}
     In this section, the local magnetic fields at the proton sites and the $^{1}$H-NMR spectrum are calculated, and the nature of the AFI phase transition is discussed.
\subsection{Model for the $^{1}$H-NMR spectra}
     Generally NMR spectra are determined by the values of static local magnetic field and the distribution of the local magnetic field at the nucleus sites in the studied material. Here, a model for the spectrum is presented and applied to $\lambda$-(BETS)$_{2}$FeCl$_{4}$. It considers all possible major sources which include the dipolar field of the Fe$^{3+}$, the exchange interactions with the Fe$^{3+}$ ion and $\pi$-electrons, the dipolar field of the neighboring proton nuclei, and the demagnetization and Lorentz contributions\cite{slichter, abragam, mori, hotta} to the local field at the proton sites.

     In $\lambda$-(BETS)$_{2}$FeCl$_{4}$ single crystals,\cite{kobayashi1} there are 16 inequivalent proton sites per unit cell (see Fig. 1). Thus, up to 16 different lines in $^{1}$H-NMR spectrum can be expected. Each of these 16 protons will have, in general, a different shift in the NMR frequency depending on its position in the crystal lattice.

     The Hamiltonian $H_{I}$ of the system for the $^{1}$H-NMR can be expressed as \cite{slichter}
\begin{equation}
  H_{I} = H_{IZ} + H_{II} + H_{Id}^{\rm{dip}} + H_{Id}^{\rm{hf}} + H_{I\pi} + H^{\rm{dem}} + H^{\rm{Lor}}, \\ 
\end{equation}
where $H_{IZ}$ is the Zeeman Hamiltonian of the $^{1}$H nuclei in $\bf{B}_{0}$, $H_{II}$ is the proton-proton nuclear dipolar interaction Hamiltonian, $H_{Id}^{\rm{dip}}$ and $H_{Id}^{\rm{hf}}$ are the dipolar coupling and transferred hyperfine coupling from the 3d Fe$^{3+}$ electrons to the protons, respectively, $H_{I\pi}$ is the hyperfine coupling of the proton nucleus to the BETS $\pi$-electrons, and the last two terms, $H^{\rm{dem}}$ and $H^{\rm{Lor}}$, are the bulk demagnetization and Lorentz contributions, respectively.\cite{slichter,carter} All of these terms contribute to the static local magnetic field at the proton sites and all but the first cause the $^{1}$H-NMR frequency shifts.

     Because of the small atomic number of the hydrogen nucleus ($Z$ = 1),\cite{slichter} it is expected that the the proton hyperfine couplings to the $\pi$-electrons ($H_{I\pi}$) and to the Fe$^{3+}$ electrons ($H_{Id}^{\rm{hf}}$) are negligible. For this reason, $H_{I}$ is simplified to
\begin{equation}
H_{I} \approx H_{IZ}+H_{II}+H_{Id}^{\rm{dip}}+H^{\rm{dem}}+H^{\rm{Lor}}.\\
\end{equation}
Among these terms it is expected that the dipolar field of the Fe$^{3+}$ ion electron spins, i.e. the contribution of Hamiltonian $H_{Id}^{\rm{dip}}$, to be the dominant source contributing to the static local magnetic field at the proton sites, as seen from the analysis in the following sections.

     Note that any interactions, both direct and indirect, between the $\pi$-electrons and the Fe$^{3+}$ ions ($\pi-$d interaction) or between the Fe$^{3+}$ ions (d$-$d interaction) will affect the polarization of the Fe$^{3+}$ electron moments, therebye modifying the dipolar field from the 3d Fe$^{3+}$ ions at the proton sites. In what follows, the effects of these interactions are considered, and the total d$-$d exchange interactions including those\cite{mori} through the Cl$^{-}$ and the conduction $\pi-$electrons (RKKY interaction) are included in the calculation using a mean field approximation to modify the Brillouin function. Since the direct $\pi-$d interaction is considered to be small because the magnetization of the conduction $\pi-$electrons is very small compared to that of the Fe$^{3+}$, it will not be included.

     The dipole moment $\vec{\mu}_{j}$ of the Fe$^{3+}$ ion $j$ produces a magnetic field $\vec{B}_{ij}$ at the proton site $i$ given by \cite{jackson}
\begin{equation}
\vec{B}_{ij} = \frac{3~\vec{r}_{ij}(\vec{\mu}\cdot\vec{r}_{ij})}{r_{ij}^{~5}} - \frac{\vec{\mu}_{j}}{r_{ij}^{~3}},\\
\end{equation}
where $\vec{r}_{ij}$ is the position vector from the proton site $i$ to the Fe$^{3+}$ ion site $j$.

     Thus, the total dipolar field $<\vec{B}_{i}>$ at the proton site $i$ is
\begin{eqnarray}
  <\vec{B}_{i}> & = &  \sum_{j}<\vec{B}_{ij}>, \\
        & = & \sum_{j}<\left[\frac{3~\vec{r}_{ij}(\vec{\mu}\cdot\vec{r}_{ij})}{r_{ij}^{~5}} - \frac{\vec{\mu}_{j}}{r_{ij}^{~3}} \right]>, \\
\nonumber
\end{eqnarray}
and the magnetization $\vec{M}(x)$ of the Fe$^{3+}$ moments is 
\begin{equation}
   \vec{M}(x) = \frac{\sum_{j}<\vec{\mu}_{j}>}{V}. \\
\end{equation}
By considering Eqs. (6)$-$(10) and (13)$-$(16), and assuming that the magnetization $\vec{M}(x)$ has the same direction as $\bf{B}$$_{0}$, one obtains
\begin{equation}
<\vec{B}_{i}> \approx\frac{g\mu_{B}J~B_{J\rm{M}}(x)}{B_{0}}\sum_{j=-N}^{+N} \left[ \frac{3\vec{r}_{ij}(\vec{B}_{0}\cdot\vec{r}_{ij})}{r_{ij}^{~5}} - \frac{\vec{B}_{0}}{r_{ij}^{~3}} \right], \\
\end{equation}
where $i$ = 1, 2, ..., 16, which indexes the proton sites, and $x$ is a $T-$dependent variable as defined by Eqs. (6)$-$(8).

     Equation (17) is used to calculate the local field at all $T$ in the PM phase. The mean field approximation for the Fe$^{3+}$ exchange interactions is included in $B_{JM}(x)$. 

     At high $T$ ($g\mu_{B}B_{0}J << k_{B}T$), Eq. (17) can also be written as
\begin{equation}
<\vec{B}_{i}> \approx\frac{(g\mu_{B})^{2}}{3k_{B}}\frac{J(J + 1)}{T + \Theta }\sum_{j=-N}^{+N} \left[ \frac{3\vec{r}_{ij}(\vec{B}_{0}\cdot\vec{r}_{ij})}{r_{ij}^{~5}} - \frac{\vec{B}_{0}}{r_{ij}^{~3}} \right] ,\\
\end{equation}
where $\Theta$ is the Curie-Weiss temperature.

     The field $\bf{B}$$_{0}$ can be expressed as $\bf{B}$$_{0}$ = $B_{0}(\sin\theta\cos\phi\hat{i} + \sin\theta\sin\phi\hat{j} + \cos\theta \hat{k}$), where $\theta$ and $\phi$ are standard spherical coordinates in the Cartesian system,\cite{arfken} and the dipolar field components $<B_{i}>_{x}$, $<B_{i}>_{y}$, and $<B_{i}>_{z}$ along x, y, z directions, respectively, can be calculated for each of the 16 inequivalent protons sites with Eq. (17) by considering the coordinates of all the inequivalent proton sites and the Fe$^{3+}$ ion positions.

     Since the values of $<\vec{B}_{i}>$ obtained in the next section obey $|<\vec{B}_{i}>|$ $<<$ $B_{0}$, the contribution of $<\vec{B}_{i}>$ to the shift of the proton spectrum comes only from the component of $<\vec{B}_{i}>$ $\parallel$ $\bf{B}$$_{0}$ ($B_{||}^{\rm{dip}}$). 
\subsection{Calculated local magnetic field at the proton sites}
\begin{figure}
\includegraphics[scale=0.41]{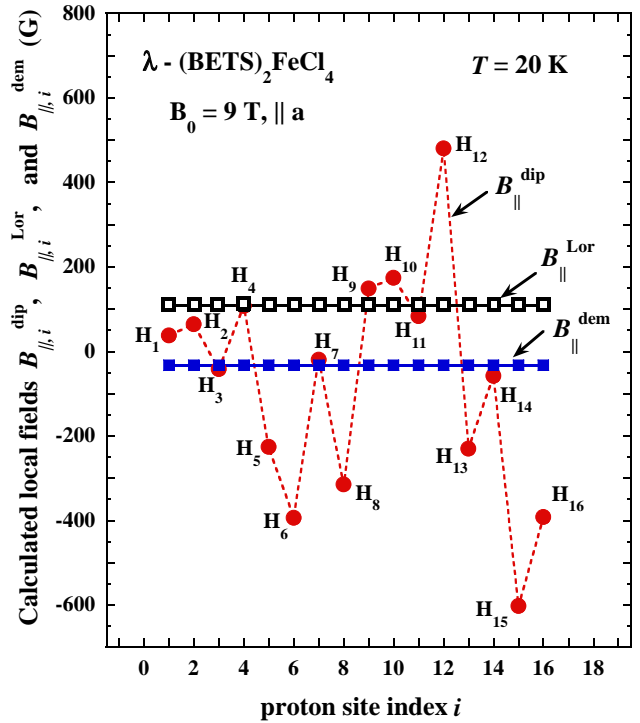}
\caption{(color online) Calculated dipolar ($B_{||}^{\rm{dip}}$), Lorentz ($B_{||}^{\rm{Lor}}$), and demagnetization field ($B_{||}^{\rm{dem}}$) components parallel to $\bf{B}$$_{0}$ at each of the 16 inequivalent proton sites (index $i$) in a single crystal of $\lambda$-(BETS)$_{2}$FeCl$_{4}$ at 20 K. The alignment of $\bf{B}$$_{0}$ = 8.9885 T is parallel to the $a$-axis in the $ac$-plane. The solid and dashed lines are guides to the eye. \label{fig5}}
\end{figure}
     Figure 5 shows the calculated component of the Fe$^{3+}$ ion dipolar field parallel to $\bf{B}$$_{0}$ at the 16 inequivalent proton sites H1, H2, ..., H16 in the crystal lattice using Eq. (17) at $T$ = 20 K and $B_{0}$ = 8.9885 T. The number of unit cells included is (2$\times$100 + 1)$^{3}$; i.e., $N$ = 100. The crystal $ac-$plane is chosen to be in the $xz-$plane in the transformation of the lattice triclinic coordinates to the Cartesian coordinates used for the calculation.

     The values of the components of the Lorentz field, $B_{||}^{\rm{Lor}}$, and the demagnetization field, $B_{||}^{\mathrm{dem}}$, parallel to $\mathbf{B}$$_{0}$, based on the shape of the needle-shape single crystal sample and the magnetization $M(x)$ [Eqs. (5)$-$(10)], are +108.3 G and $-$32.5 G, respectively,\cite{lorentz, carter} as shown in Fig. 5. The net shift from both of these contributions is their sum, i.e. +75.8 G, which is small, but not completely negligible. The small spacial variations\cite{carter} of the demagnetization field across the sample have been neglected.

     Also, the calculated contribution of $H_{II}$ is only $\leq$ 3 G among the 16 inequivalent proton sites, as confirmed by the spin-echo decay measurements.\cite{wu} Since it is so small, it is also neglected here.

     Thus, $B_{||}^{\mathrm{dip}}$ obtained from Eq. (17) (see Fig. 5) is the dominant contribution to the structure of $\chi^{\prime\prime}$($\nu$). Since $B_{||}^{\mathrm{Lor}}$ and $B_{||}^{\mathrm{dem}}$ are nearly constant over all the proton sites, they have a negligible effect on the structure of $\chi^{\prime\prime}$($\nu$); their contribution constitutes a shift in frequency or field but with almost the same amount for each proton sites ($\sim$ 25$\%$ to the average $\Delta\nu$ at 20 K).

     The values of $B_{||}^{\mathrm{dip}}$ at 20 K cover a wide range of field, from $\pm$ 20 G up to $\pm$ 600 G, depending on the proton positions in the crystal lattice. For example, at the proton site H15 it is $\sim$ $-$600 G, and $\sim$ +500 G at the proton site H12. The positions of these proton sites are all labeled in Fig. 1. The proton site (H15) that has the largest $B_{||}^{\mathrm{dip}}$ from the Fe$^{3+}$ ions is that which is closest to the nearby Fe$^{3+}$ ion plane in the crystal lattice. The range of field ($\sim$ $-$600 $-$ +600 G) that $B_{||}^{\mathrm{dip}}$ covers corresponds to a range of frequency of $\sim$ 5 MHz. 
\subsection{Calculated $^{1}$H-NMR spectra from the dipolar field contributions}
\begin{figure}
\includegraphics[scale= 0.65]{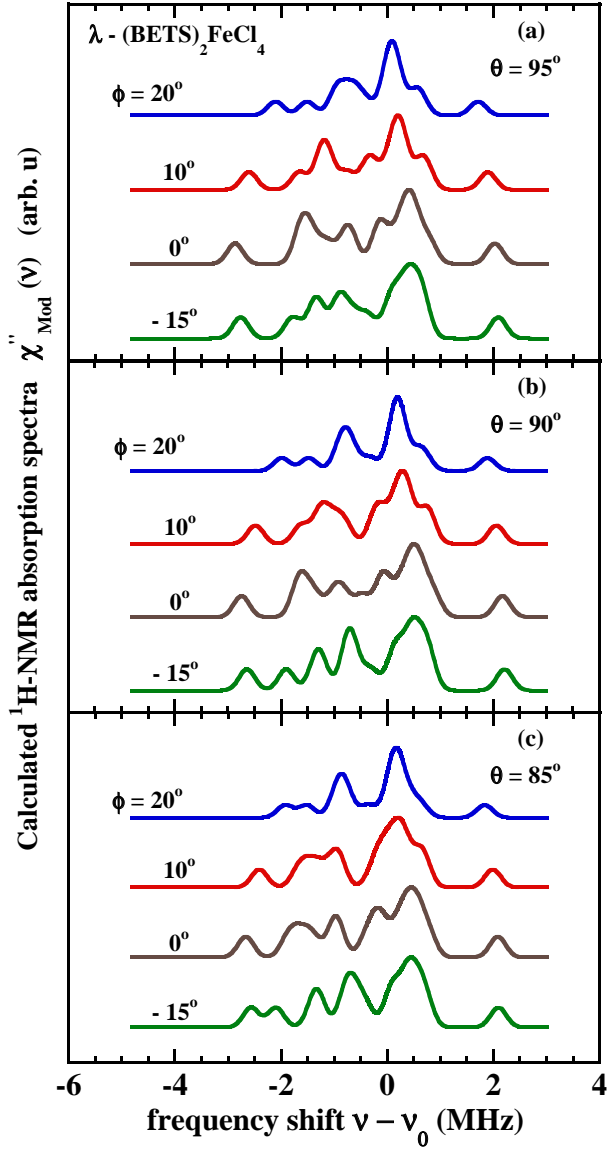}
\caption{(color online) Calculated proton $\chi_{\rm{mod}}^{''}(\nu)$ from the 3d Fe$^{3+}$ ion electron dipolar contributions in single crystal $\lambda$-(BETS)$_{2}$FeCl$_{4}$ at 20 K with $\bf{B}$$_{0}$ = 8.9885 T aligned close to the $a$-axis ($\theta$ = 90$^{\circ}$ $\pm$ 5$^{\circ}$) and near the $ac$-plane ($-$15$^{\circ}$ $\leq$ $\phi$ $\leq$ +20$^{\circ}$). The Larmor frequency $\nu_{0}$ = 382.6935 MHz. \label{fig6}}
\end{figure} 
     In this section the model for the proton absorption spectrum $[\chi_{\rm{mod}}^{''}(\nu)]$ is calculated. The first step is to calculate the static local magnetic field $B_{||,i}^{\rm{dip}}$ at each of the 16 inequivalent proton sites, as described in Sections IV A and B. The second step is to convolve this field distribution with a set of Gaussian functions, $y_{i}$, each of which has a maximum amplitude of 1 and the same width $\delta$ at each proton site $i$.\cite{abragam} In this case,
\begin{equation}
\chi_{\mathrm{mod}}^{''}(\nu) = \sum_{i=1}^{16}\exp\left[ -\frac{(\nu- \gamma_{0} B_{||,i}^{\rm{dip}})^{2}}{2\delta^{2}} \right] .\\
\end{equation}

      Plots of $\chi_{\mathrm{mod}}^{''}(\nu)$, calculated with this model and the value of $\delta$ = 0.15 MHz, are shown in Fig. 6 for $T$ = 20 K with $B_{0}$ = 8.9885 T aligned close to the $a$-axis ($\theta$ = 90$^{\circ}$ $\pm$ 5$^{\circ}$) and near the $ac$-plane ($-15^{\circ} \leq \phi \leq +20^{\circ}$). These angles are selected because they are close to what is needed for comparison with the measurements. As can be seen in Fig. 6, $\chi_{\rm{mod}}^{''}(\nu)$ is fairly sensitive to $\theta$ and $\phi$ which are determined by the direction of $\bf{B}$$_{0}$.
\subsection{Comparison of the model to the measured spectra}
      Figure 7 shows a comparison of the calculated $\chi_{\rm{mod}}^{''}(\nu)$ (upper) using $B_{0}$ = 8.9885 T with $\theta$ = +85$^{\circ}$ and $\phi$ = +5$^{\circ}$, and the measured $\chi^{''}(\nu)$ (lower) at 20 K for $\nu_{0}$ = 382.635 MHz. The calculated result includes the dipolar field from the Fe$^{3+}$ electrons, and corrections for the Lorentz and bulk demagnetization fields. 
\begin{figure} 
\includegraphics[scale= 0.42]{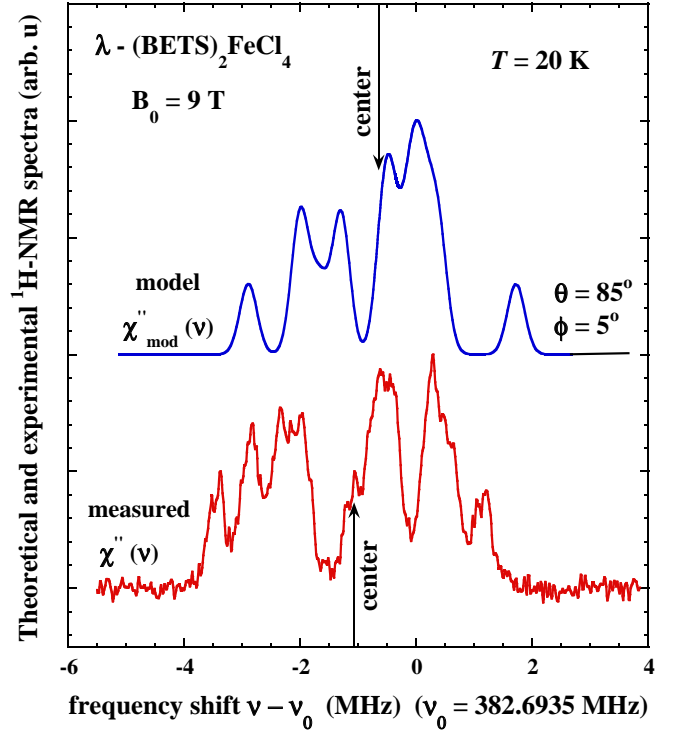} 
\caption{(color online) Comparison of the calculated $\chi_{\rm{mod}}^{''}(\nu)$ $[$upper (blue) smooth line$]$ with the measured $^{1}$H-NMR absorption spectrum $\chi^{''}(\nu)$ $[$lower (red) line$]$ in a single crystal of $\lambda$-(BETS)$_{2}$FeCl$_{4}$ at 20 K. For $\chi_{\rm{mod}}^{''}(\nu)$, the value of $\bf{B}$$_{0}$ is 8.9885 T aligned with $\theta$ = +85$^{\circ}$ and $\phi$ = +5$^{\circ}$, which is close to the $a$-axis and near the $ac$-plane. The vertical arrows indicate the center (first moment $<\nu>$) of each line. \label{fig7}}
\end{figure}

      There are several small differences between the measured and calculated model spectra shown in Fig. 7. One is that at 20 K, the shift in the center of the spectrum differs by $\sim$ 42 kHz ($\sim$ 100 G). This is quite small compared to the separation of the outer peaks, which is 4.65 MHz (1.09 kG) for both the measured and calculated spectra. On the other hand, $\Delta f_{\rm{rms}}$ = 1.126 MHz and 1.387 MHz for the model and measured values, respectively, corresponding to a difference of $\sim$ 18$\%$. Another small difference is in the shape of the spectra. These similarities in the model and measured spectra strongly support the conclusion that the proton spectrum in $\lambda$-(BETS)$_{2}$FeCl$_{4}$ is dominated by the dipole field of the Fe$^{3+}$ ions.

     An important question is what is responsible for the small difference between the spectra calculated with the model that has been used and the measured spectra. It is probably caused by not including all the details of the exchange interactions (including the d$-$d exchange interactions through the Cl$^{-}$ and through the BETS conduction $\pi-$electrons) between the Fe$^{3+}$ electron spins when using the mean field $M(x)$ or $B_{J\rm{M}}(x)$ to model the Fe$^{3+}$ electron spin polarization.

     Also, an important aspect of the $\pi-$d interaction which is also part of the d$-$d exchange interactions is that it should be responsible for the coordination of the occurrence of the PM$-$AFI phase transition.

     It is expected that if all of the exchange interactions are included in detail, they will cause a small change in the moment of the Fe$^{3+}$ ions and their polarization will not be precisely along $\bf{{B}_{0}}$, even in the PM phase. It is likely that these additional factors are responsible for the relatively small difference between the measured and calculated shift and width in the $^{1}$H-NMR spectrum. To include them in a model is beyond the scope of this paper. 
\subsection{The local magnetic field in the AFI state and the nature of the PM$-$AFI phase transition}
     As in the PM state, the local magnetic field at the proton sites in the AFI state ($T$ $\leq$ $T_{MI}$ = 3.5 K) is also dominated by the dipolar field from the Fe$^{3+}$ ions, even though Eq. (17) is not specific to and may not apply in the AFI state. In both phases, the Fe$^{3+}$ ion electron spin moments are present and the general Hamiltonian $H_{I}$ of the $^{1}$H-NMR system Eq. (11) or (12) applies in both the PM and AFI states.

     However, the change of the $^{1}$H-NMR spectra in $\lambda$-(BETS)$_{2}$FeCl$_{4}$ is significant. On cooling from the PM to the AFI state, the spectrum broadens, the splittings are smeared out, and a new peak appears on the high frequency side.

     It is unlikely that these changes are caused by variations in the Lorentz or demagnetization field because the contribution from each of them is essentially the same at each proton site, even though they are proportional to the Fe$^{3+}$ ion electron susceptibility.

     Therefore the changes of the details in the proton spectra at any $T$ (including in the PM and AFI phases) come mainly from the change of the dipolar field of the 3d Fe$^{3+}$ ions, which in part are changed by the effect of $\pi-$d and d$-$d interactions on the polarization of the Fe$^{3+}$ moments. The major difference for the Fe$^{3+}$ moments is that they should have long-range order in the AFI state, which is formed by the $\pi-$d and d$-$d interactions. \cite{uji1,akutsu2, akutsu3}

     The evidence from the change of the $^{1}$H-NMR spectra, as well as that reflected by the discontinuities of the frequency shift $\Delta\nu$ and the rms linewidth $\Delta f_{\mathrm{rms}}$, is indicative of a first order nature for the PM$-$AFI phase transition in $\lambda$-(BETS)$_{2}$FeCl$_{4}$. \cite{mori, watanabe}
\subsection{Comparison with other results}
     In this section, we first compare our results on a single $\sim$ 4 $\mu$g crystal with those reported by Endo et al.\cite{endo} on a large ($\sim$ 6.5 mg) aggregate of crystals aligned along the $c$-axis. One major difference is that since their $\bf{B}$$_{0}$ = 2.2 T was applied perpendicular to the $c$-axis, the corresponding $^{1}$H-NMR spectrum is the average over all alignments of $\bf{B}$$_{0}$ perpendicular to the $c$-axis. It is expected that in comparison to the measurements reported here on a single crystal, this average will smear out some of the details of the spectrum and will generate a broader range of Lorentz and demagnetization fields that also have an effect on the spectrum. Another important difference is that because $\bf{B}$$_{0}$ in our measurements is 4.1 times larger than that used in their work, the splittings and shifts of the spectral features is expected to be 4.1 times larger; i.e., the spectral resolution of our 9 T measurements is significantly higher.

     Their results \cite{endo} showed the onset of a splitting of the spectrum near 70 K that was considered an anomaly and interpreted as possible evidence for a charge disproportionation of the $\pi$-electrons associated with a ferroelectric-type phase transition. They further suggested \cite{endo} that this splitting is from the hyperfine field between the proton nucleus and $\pi$ electrons and not the dipole field of the Fe$^{3+}$ ions. 

     Our results disagree with their results and interpretation in several ways. First, our measurements (Fig. 2) show a continuous increase in the splitting and the shift (Fig. 3) of the spectrum as $T$ is decreased from 180 to $\sim$ 5 K. No evidence for an anomaly near 70 K is observed and the main features of the spectra are well explained in terms of the model based upon the dipole field of the Fe$^{3+}$ ions.

     Another problem with the interpretation by Endo et al.\cite{endo} is that if the spectral splitting is caused by the hyperfine field between the proton nucleus and $\pi$ electrons, it would require a large hyperfine shift of $\sim$ 50 G, corresponding to a Knight shift of $\sim$ 0.23$\%$. It is very unlikely that such a large shift could occur for the very light nucleus of a hydrogen atom.

     Experimetal evidence that such a large hyperfine field does not occur at the proton sites is also indicated by comparing the proton spin-lattice relaxation rate (1/$T_{\text{1}}$) at 100 K in\cite{wgc} $\lambda$-(BETS)$_{2}$FeCl$_{4}$ and\cite{takagi} $\lambda$-(BETS)$_{2}$GaCl$_{4}$. Both materials have the same structure and related properties, but without the magnetization of the Fe$^{\text{3+}}$ ions in the Ga compound. If the main fluctuating magnetic field at the proton site were from the $\pi$ electrons, one would expect approximately the same value of 1/$T_{\text{1}}$ for both materials. Instead, at 100 K 1/$T_{\text{1}}$ has a value that is $\sim$ 10$^{3}$ times larger in $\lambda$-(BETS)$_{2}$FeCl$_{4}$ than in $\lambda$-(BETS)$_{2}$GaCl$_{4}$. Futhermore, in $\lambda$-(BETS)$_{2}$FeCl$_{4}$, 1/$T_{1}$ has the $T-$dependence expected from the saturation of the magnetization of the Fe$^{3+}$ ions. \cite{wgc}

     These properties provide strong support to our interpretation that the dominant internal magnetic field at the protons in $\lambda$-(BETS)$_{2}$FeCl$_{4}$ is the dipole field of the Fe$^{\text{3+}}$ ions, and that there is no anomalous splitting of the spectrum near 70 K.

     Other interpretations of a ferroelectric phase transition at 70 K in $\lambda$-(BETS)$_{2}$FeCl$_{4}$ are based upon the discontinuity reported for its specific heat,\cite{negishi} the division of the electron spin resonance $g-$factor from one into two values,\cite{brossard} and the increase in the microwave dielectric constant.\cite{matsui} It is not yet clear why the effect of such a transition on the magnetization of the Fe$^{3+}$ ions is so small that it is not seen in our $^{1}$H-NMR spectrum measurements.
\section{Conclusions}
     $^{1}$H-NMR spectrum measurements on a small ($\sim$ 4 $\mu$g) single crystal of the organic conductor $\lambda$-(BETS)$_{2}$FeCl$_{4}$ using an applied magnetic field $B_{0}$ = 8.9885 T parallel to the $a$-axis in the $ac$-plane over the temperature range 2.0$-$180 K are reported. This work, and a preliminary report of it,\cite{wgc} are the first NMR reports that use a single crystal of $\lambda$-(BETS)$_{2}$FeCl$_{4}$. The results provide the distribution of the static local magnetic field at the proton sites in both the PM and the AFI phases.

     The experimental spectra have six main peaks and become progressively broadened and shifted as $T$ is decreased from 180 K to $\sim$ 5 K. For $T$ $\leq$ $T_{MI}$ = 3.5 K (below the PM$-$AFI transition), an extra peak appears on the high frequency side, the details of the spectrum become smeared, and changes in the frequency shift and the rms linewidth are discontinuous, indicating a significant change in the static local magnetic field distribution at the proton sites on traversing the PM to AFI phase transition. 

     The origin of the spectral features is attributed to the large dipolar field from the 3d Fe$^{3+}$ electron moments (spin $S_{d}$ = 5/2, $g$ $\approx$ 2) at the proton sites. The main features of the spectra are successfully modeled with a mean field corrected Brillouin function. 

     The value for $J_{\rm{dd}}$ between each of the two nearest neighbor Fe$^{3+}$ ions obtained from this fit to the measurements is $\sim$ $-$ 0.85 K, which is close to the theoretical prediction.\cite{mori}

     No NMR evidence for an anomaly at 70 K reported earlier\cite{endo} on an aggregate of crystals is observed.
 
     It is suggested that the smaller features of the spectra that are not covered by this model are caused by the electron-electron interactions that are beyond the scope of this paper. 
\begin{acknowledgments}
     This work is supported at UCLA by NSF Grant DMR$-$0334869 (WGC) and 0520552 (SEB), partial support at NHMFL by NSF under cooperative agreement DMR$-$0084173, and that at Indiana by Petroleum Research fund ACS-PRF 33912-AC1. We thank A. Kobayashi and H. Kobayashi for the crystal structure data, and thank G. Gaidos, F. Zamborszky, J. Shinagawa, and F. Zhang for helpful discussions. 
\end{acknowledgments}

\begin{references}
%
\bibitem{uji1} S. Uji, H. Kobayashi, L. Balicas, and J. S. Brooks, Adv. Mater. $\bf{14}$, 243 (2002).

\bibitem{uji2} S. Uji, H. Shinagawa, T. Terashima, T. Yakabe, Y. Terai, M. Tokumoto, A. Kobayashi, H. Tanaka, and H. Kobayashi, Nature $\bf{410}$, 908 (2001).

\bibitem{tokumoto} M. Tokumoto, T. Naito, H. Kobayashi, A. Kobayashi, V. N. Laukhin, L. Brossard, and P. Cassoux, Synth. Met. $\bf{86}$, 2161 (1997).

\bibitem{brossard} L. Brossard, R. Clerac, C. Coulon, M. Tokumoto, T. Ziman, D. K. Petrov, V. N. Laukhin, M. J. Naughton, A. Audouard, F. Goze, A. Kobayashi, H. Kobayashi, and P. Cassoux, Eur. Phys. J. B $\bf{1}$, 439 (1998).

\bibitem{kobayashi1} H. Kobayashi, H. Tomita, T. Naito, A. Kobayashi, F. Sakai, T. Watanabe, and P. Cassoux, J. Am. Chem. Soc. $\bf{118}$, 368 (1996).

\bibitem{akutsu1} H. Akutsu, E. Arai, H. Kobayashi, H. Tanaka, A. Kobayashi, and P. Cassoux, J. Am. Chem. Soc. $\bf{119}$, 12681 (1997).

\bibitem{endo} S. Endo, T. Goto, T. Fukase, H. Matsui, H. Uozaki, H. Tsuchiya, E. Negishi, Y. Ishizaki, Y. Abe, and N. Toyota, J. Phys. Soc. Jpn. $\bf{71}$, 732 (2002).

\bibitem{negishi} E. Negishi, H. Uozaki, Y. Ishizaki, H. Tsuchiya, S. Endo, Y. Abe, H. Matsui, and N. Toyota, Synthetic Metals $\bf{133-134}$, 555 (2003).

\bibitem{matsui} H. Matsui, H. Tsuchiya, T. Suzuki, E. Negishi, and N. Toyota, Phys. Rev. B $\bf{68}$, 155105 (2003).

\bibitem{jaccarino} V. Jaccarino and M. Peter, Phys. Rev. Lett. $\bf{9}$, 290 (1962).

\bibitem{balicas1} L. Balicas, J. S. Brooks, K. Storr, S. Uji, M. Tokumoto, H. Tanaka, H. Kobayashi, A. Kobayashi, V. Barzykin, and L. P. Gor$'$kov, Phys. Rev. Lett. $\bf{87}$, 067002 (2001).

\bibitem{balicas2} L. Balicas, V. Barzykin, K. Storr, J. S. Brooks, M. Tokumoto, S. Uji, H. Tanaka, H. Kobayashi, and A. Kobayashi, Phys. Rev. B $\bf{70}$, 092508 (2004).

\bibitem{fulde} P. Fulde and R. A. Ferrell, Phys. Rev. A $\bf{135}$, 550 (1964); A. I. Larkin and Yu. N. Ovchinnikov, Sov. Phys. JETP $\bf{20}$, 762 (1965).

\bibitem{houzet} M. Houzet, A. Buzdin, L. Bulaevskii, and M. Maley, Phys. Rev. Lett. $\bf{88}$, 227001 (2002); L. N. Bulaevskii, Sov. Phys. JETP $\bf{38}$, 634 (1974).

\bibitem{akutsu2} H. Akutsu, K. Kato, E. Ojima, H. Kobayashi, H. Tanaka, A. Kobayashi, and P. Cassoux, Phys. Rev. B $\bf{58}$, 9294 (1998).

\bibitem{akutsu3} H. Akutsu, K. Kato, E. Arai, H. Kobayashi, A. Kobayashi, M. Tokumoto, L. Brossard, and P. Cassoux, Solid State Communications $\bf{105}$, 485 (1998).

\bibitem{kobayashi2} H. Kobayashi, H. Tomita, T. Udagawa, T. Naito, and A. Kobayashi, Synthetic Metals $\bf{70}$, 867 (1995).

\bibitem{kobayashi3} H. Kobayashi, A. Kobayashi, F. Sakai, and P. Cassoux, Chem. Soc. Rev. $\bf{29}$, 325 (2000).

\bibitem{wgc} W. G. Clark, Guoqing Wu, P. Ranin, L. K. Montgomery, and L. Balicas, Appl. Magn. Reson. $\bf{27}$, 279 (2004).

\bibitem{hk} H. Kobayashi, E. Fujiwara, H. Fujiwara, H. Tanaka, H. Akutsu, I. Tamura, T. Otsuka, A. Kobayashi, M. Tokumoto and P. Cassoux, J. Phys. Chem. Solids $\bf{63}$, 1235 (2002).

\bibitem{montgomery} L. K. Montgomery, T. Burgin, T. Miebach, D. Dunham, and J. C. Huffman, Mol. Cryst. Liq. Cryst. $\bf{284}$, 73 (1996).

\bibitem{clark} W. G. Clark, M.E. Hanson, F. Lefloch, and P. S$\acute{e}$gransan, Rev. Sci. Instrum. $\bf{66}$, 2453 (1995).

\bibitem{slichter} C. P. Slichter, $\it{Principles ~of~ Magnetic~ Resonance}$ (Springer, Berlin, 1989), 3rd ed..

\bibitem{abragam} A. Abragam, $\it{The~ Principles~ of~ Nuclear~ Magnetism}$ (Claredon Press, Oxford, 1962).

\bibitem{mori} T. Mori and M. Katsuhara,  J. Phys. Soc. Jpn. $\bf{71}$, 826 (2002).

\bibitem{ashcroft} N. W. Ashcroft and N. D. Mermin, $\it{Solid ~State~ Physics}$ (Holt, Rinehart and Winston, New York, 1976), 1st ed..

\bibitem{hotta} C. Hotta and H. Fukuyama, J. Phys. Soc. Jpn. $\bf{69}$, 2577 (2000).

\bibitem{carter} G. C. Carter, L. H. Bennett, and D. J. Kahan, $\it{Metallic ~Shifts ~in ~NMR}$ (Pergamon, London, 1977), part I.

\bibitem{jackson} J. D. Jackson, $\it{Classical ~Electrodynamics}$ (John Wiley $\&$ Sons, Singapore, 1990), 2rd ed..

\bibitem{arfken} G. B. Arfken and H. J. Weber, $\it{Mathematical~ Methods~ for~ Physicists}$ (Academic Press, Inc., San Diego), 4th ed..

\bibitem{lorentz} The Lorentz field $B^{\rm{Lor}}$ and demagnetization field $B^{\rm{dem}}$ in $\lambda$-(BETS)$_{2}$FeCl$_{4}$ can be expressed respectively as, $B^{\rm{Lor}}$ = $\frac{4\pi}{3} \frac{M(x)}{N_{A}\upsilon_{Fe}}$ and $B^{\rm{dem}}$ = 4$\pi D \frac{M(x)}{N_{A}\upsilon_{Fe}}$, where $D$ is the demagnetization factor depending on sample size, $M(x)$ is the magnetization of the 3d Fe$^{3+}$ electrons, $N_{A}$ is the Avogadro number, $\upsilon_{Fe}$ is the unit cell volume per Fe$^{3+}$ ion, and $B_{0}$ is the applied magnetic field. Here $D$ $\approx$ 0.1 according to the size of the needle shape single crystal.

\bibitem{wu} Guoqing Wu, P. Ranin, and W. G. Clark, unpublished.

\bibitem{watanabe} M. Watanabe, S. Komiyama, R. Kiyanagi, Y. Noda, E. Negishi, and N. Toyata, J. Phys. Soc. Jpn. $\bf{72}$, 452 (2003).

\bibitem{takagi} S. Takagi et al., J. Phys. Soc. Jpn. $\bf{72}$, 3259 (2003).
%
\end{references}
\end{document}